\documentclass[runningheads]{llncs}

\usepackage{graphicx}
\usepackage{url}
\usepackage{xcolor}
\usepackage{soul}
\usepackage{tikz}
\usepackage{hyperref}

\begin{document}

\title{
Rethinking Programming Paradigms\\
in the QC-HPC Context\thanks{\tiny This manuscript has been authored by UT-Battelle, LLC, under contract DE-AC05-00OR22725 with the US Department of Energy (DOE). The US government retains a nonexclusive, paid-up, irrevocable, worldwide license to publish or reproduce the published form of this manuscript, or allow others to do so, for US government purposes. DOE will provide public access to these results of federally sponsored research in accordance with the \href{DOE Public Access Plan}{https://www.energy.gov/doe-public-access-plan}. This is the author’s version of the work. It is posted here for your personal use. Not for redistribution. The definitive version was published in Springer Nature at \url{https://doi.org/10.1007/978-3-031-61763-8_8}.}
}
\titlerunning{QC-HPC}

\author{
Silvina Caino-Lores\inst{3}
\and
Daniel Claudino\inst{1}
\and
Eugene Dumitrescu\inst{1}
\and \\
Travis S.~Humble\inst{1}
\and
Sonia Lopez Alarcon\inst{2}
\and 
Elaine Wong\thanks{\tiny Corresponding Author}\inst{1}
} 
\authorrunning{Wong et al.}

\institute{
Oak Ridge National Laboratory, Oak Ridge TN, USA\\
\email{\{claudinodc,dumitrescuef,humblets,wongey\}@ornl.gov} \and
Rochester Institute of Technology, Rochester NY, USA\\
\email{slaeec@rit.edu} \and
University of Rennes, Inria, CNRS, IRISA; Rennes, France \\
\email{silvina.caino-lores@inria.fr}
}

\maketitle

\begin{abstract}
Programming for today’s quantum computers is making significant strides toward modern workflows compatible with high performance computing (HPC), but fundamental challenges still remain in the integration of these vastly different technologies. Quantum computing (QC) programming languages share some common ground, as well as their emerging runtimes and algorithmic modalities. In this short paper, we explore avenues of refinement for the quantum processing unit (QPU) in the context of many-tasks management, asynchronous or otherwise, in order to understand the value it can play in linking QC with HPC. Through examples, we illustrate how its potential for scientific discovery might be realized.
\end{abstract}

\section{Introduction}

A quantum computer is a physical system in which an initial quantum state evolves in time quantum mechanically to reach a new state that will contain the solution to a certain problem. Like other accelerators such as GPUs, ASICs and FPGAs, quantum computers have their own niche applications at which they excel, and are not meant to universally outperform classical computing. The set of problems that will benefit from quantum computing (QC) is still being defined, but some research points to Hamiltonian simulation as a good first candidate, which has practical applications that will enhance drug development~\cite{Chen2024}. QC has also shown advantages over classical implementations in very specific, but arguably not useful problems~\cite{Arute2019,Madsen2022}. More optimistically, however, research on all aspects of QC from the technology to the software stack and high performance computing (HPC) integration is advancing the field to reach quantum advantage on real life problems.

QC is intricate in its workings, but as the software stack advances, and applications are defined, non-quantum experts should be able to take advantage of these accelerators without deep knowledge of quantum entanglement, quantum superposition, or quantum noise mitigation \cite{LopezAlarcon2022,Schulz2022,LopezAlarcon2023}. One aspect that users will need to be aware of is the \textit{probabilistic nature of quantum mechanics}. The quantum state of a quantum system cannot be observed by classical means. The quantum state can be in a superstition of waves similar to the way sound can contain waves on different frequencies. But in quantum mechanics, a classical observation can only see one of these frequencies, and the others will collapse and be un-observable. For that reason, the measurement is done by repeated observation, through which each specific frequency or wave is observed with a certain probability. The building of this probability distribution contains information that can be used to solve computational problems. Therefore, quantum measurement requires repeated stochastic runs (shots) and measurements to extract the quantum state's probability distribution on a certain quantum basis.

Accelerated computing has become a leading architectural approach for HPC systems. This paper explores the possibility of quantum computers claiming their role as hardware accelerators, with the understanding that their architectures rely on the presence of specialized devices to process specialized workloads or tasks. The ways that this can be combined effectively remains to be seen and for the moment, a concrete realization of how to integrate a quantum computer with a HPC system would not necessarily demonstrate a performance advantage with current systems, but would be able to show that testing such an integration is \textit{feasible with current programming tools}. In this paper, we outline how computation tasks might interact with each other in this context on known quantum programs.

\section{Quantum Programming Tools}

\begin{figure*}[ht]
\centering
\includegraphics[scale=0.25]{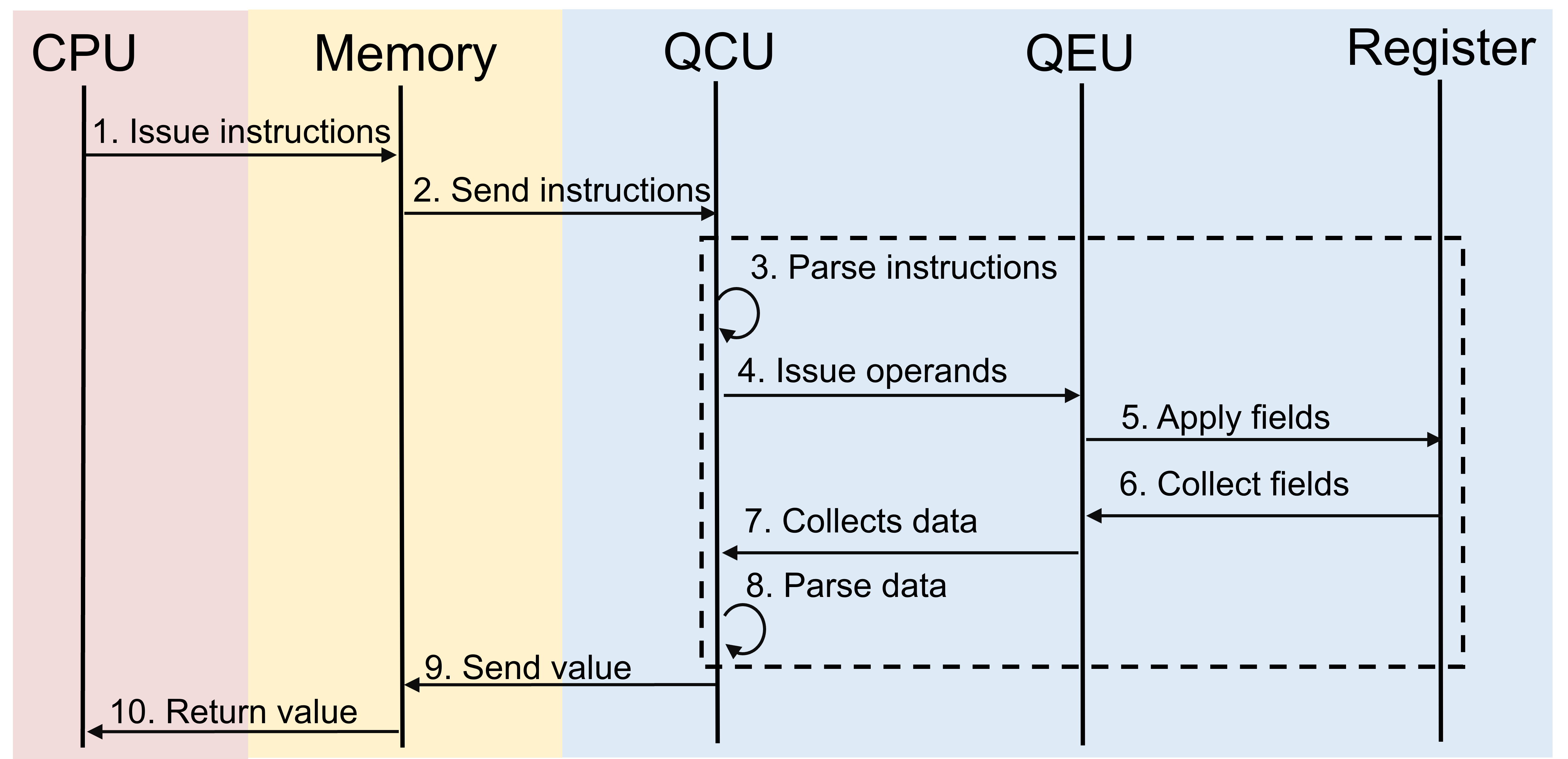}
\label{fig:nodeexecute}
\caption{A sequence diagram detailing the interactions between a CPU, memory, and a QPU~\cite{Britt2017}. Operands are the sequence of quantum gates that implement an input instruction logic, the quantum control unit (QCU) parses these instructions while the quantum execution unit (QEU) initiates the corresponding gates through applied fields that modify the quantum physical state of the quantum register. The return process generates data that corresponds with measured results from hardware.}
\end{figure*}

Despite the nascent technology underlying QC, there has been substantial progress in the development of programming tools that integrate QPUs with conventional systems. Issuing instructions to a quantum control unit plays a pivotal role in the execution of a quantum program, while the recovering of data from the QPU depends on subsequent post-processing of measurement results. This interplay between the QPU and the calling system--depicted in Fig.~\ref{fig:nodeexecute}--can be developed to be either synchronous or asynchronous depending on the algorithm, but the latter is more generally used at the moment.

One way of classifying programming paradigms involves its distance from the program to the hardware. Thus, how tasks are to be viewed might be relative to how easy it is to access hardware and manage its latency, resources, and general effectiveness. At the analog level, pulse sequences drive the transformation of a quantum state, circumventing the need to involve the full stack in computation. At the digital level, the instructions to manage computations can range from low-level (close to hardware) assembly languages, which can be transformed to signals for physical implementation, to a higher level that can support algorithmic descriptions closer to the user's natural language, and then brought down closer to hardware via a compiler. Recent integration attempts~\cite{Mccaskey2020,Mintz2020} were motivated by the desire to balance ease-of-use for the general user and flexibility for the domain scientists, but those frameworks and paradigms still sit quite close to the hardware without a consensus on how to approach quantum high performance computing (QHPC) task management at a system level.

\begin{figure*}[ht]
\centering
\includegraphics[scale=0.5]{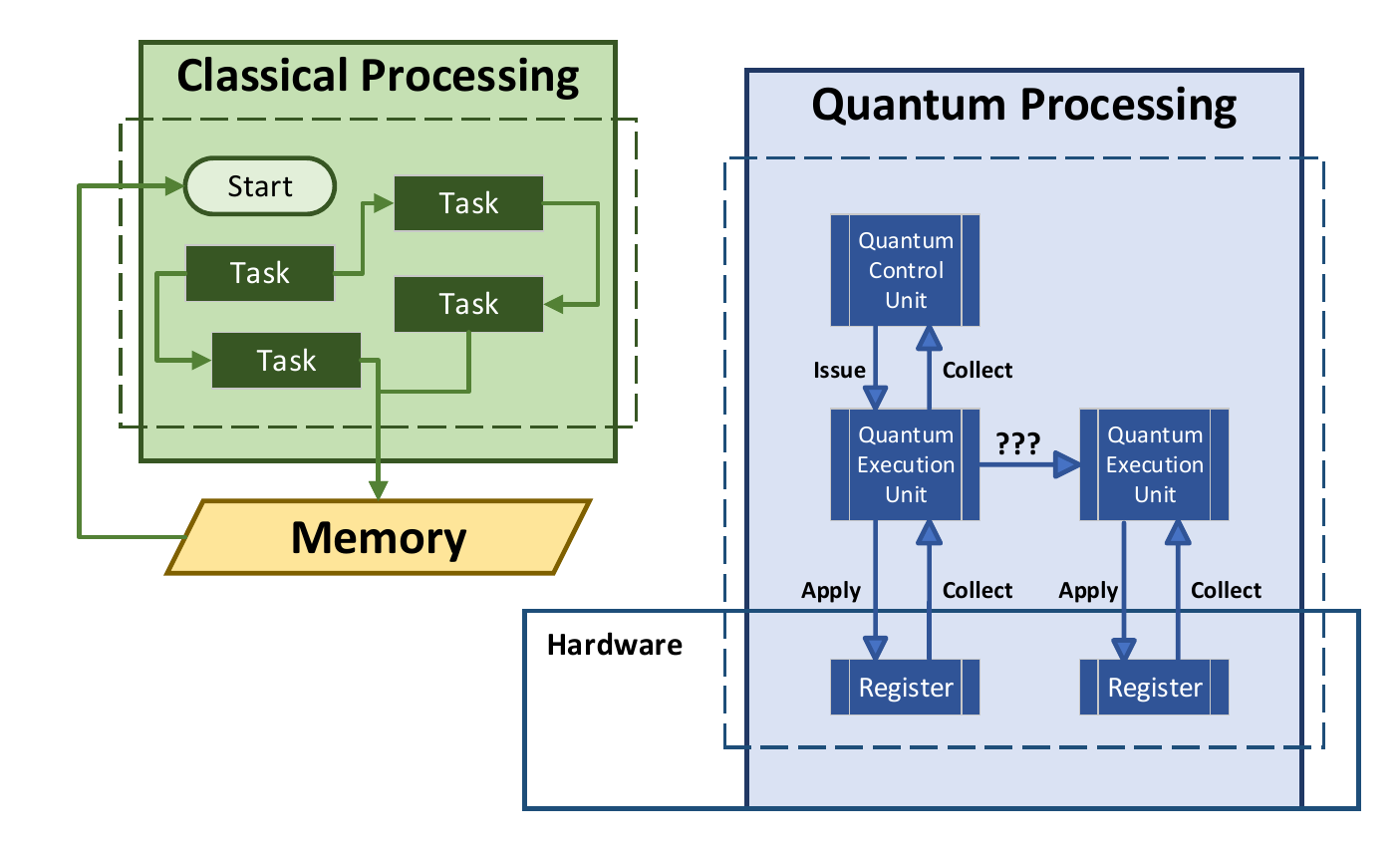}
\vspace{-5pt}
\label{fig:qshighlevel}
\caption{An example of a high-level view of the task-based interactions in a hybrid quantum-classical system.} 
\end{figure*}

We begin to consider a higher-level perspective revolving around the idea of runtimes, task-oriented and workflow-oriented approaches in QC and HPC integration for the purpose of \textit{interoperability}. An example of this is shown in Fig.~\ref{fig:qshighlevel}. 
This might be most akin to hybrid runtime environments with multi-processing, multi-threading and accelerator offloading capabilities (e.g., MPI, OpenMP and CUDA, respectively), where similar challenges inherent in the handling of heterogeneous systems arise: there are trade-offs with respect to the overhead of offloading a computation to an accelerator vs. the obtained performance improvement. A vast amount of literature exists on this topic, with years of experience and incremental optimizations. But, while popular solutions in HPC exist, there is a shift to try to find more flexible, productive, and higher-level mechanisms to define tasks, dependencies, resource requirements and other constraints, especially with systems that operate different physical systems from traditional computing platforms.

For example, let's consider a machine learning model implemented in TensorFlow running on CUDA/OpenACC vs. an AI-powered workflow doing ensemble molecular dynamics at scale through a true workflow management system (e.g., PyCOMPSs). In this scenario, the level of integration moves away from the interaction between the accelerator and the CPU to involve multiple nodes potentially dedicated to different types of tasks. HPC systems are remotely accessible with resources assigned for a particular task via a scheduler and resource manager. In QC, existing approaches for this kind of system-level integration usually go through similar remote access by necessity, as there are strict environmental constraints for operating a quantum system. However, such system integration is not tightly-coupled and would not be managed as in HPC, but is nevertheless a tangible scenario for actual workflow composition, and we already see solutions that allow writing hybrid quantum-classical workflows running on cloud-based QC-HPC systems (e.g., Covalent, which supports AWS Braket and IBM Quantum on the QC side).

We can begin to understand the interoperability challenges of a hybrid application use case by analyzing it holistically in a top-down manner (i.e., starting from the high-level workflow and finishing at the low-level programming of the hardware instructions). For example, one might consider the following progression of analysis:
\begin{enumerate}
\item Identify the high level steps.
\item Identify, out of those steps, which ones will be suitable for the classical side and which ones will be suitable for the quantum side. For the parts that can be accelerated by quantum computers, one needs to understand the existing quantum algorithm and how to encode the information. Such translations might be easier for physics problems. Since the applications that benefit from QC acceleration are not well-defined yet, understanding how information is handled in the quantum world becomes essential to make use of this information on the classical world. The desired approach is that this quantum algorithm and its quantum information encoding can be abstracted away from the user.
\item Define how information is going to be exchanged between the classical and quantum sides, which would require some sort of memory at both ends to account for the different representations of data on each side. Furthermore, CPU and GPU can share memory but CPU and QPU cannot. This encoding-decoding from classical to quantum and vice-versa is a challenging and computationally intensive step.
\item Identify existing integration tools with access to possibly several backends. It will be necessary to define the call to quantum kernels and what those kernels will do depending on the compiler requirements. The runtime or workflow manager would account for latency, memory management, and task interactions, while the compiler takes care of generating a synthesizable quantum instruction stream from a high level quantum program description.
\end{enumerate}

\section{Task Modeling in Quantum Computation}\label{sec:examples}

This section frames quantum programs in terms of their related tasks and where asynchronous multi-tasking requirements might appear in the context of a hybrid workflow management. However, we note that existing QC hardware is suitable for both synchronous and asynchronous operations, and we provide \textbf{two examples} to illustrate this. The distinction here represents a choice for how the QPU behaves in response to instructions accepted from the parent node within the accelerator architecture. This choice depends on the desired behavior of the QPU and, for purposes of performance, on the ability for the QPU to process an assigned workload in a well-characterized time. For example, synchronous operations of a QPU place a timing constraint on the compute response that depends intimately on the size of the quantum program, i.e., the depth of the circuit, and the number of measurements samples generated through repeated execution of the program. While these parameters are well-defined prior to execution, the subsequent impact on the execution time varies with the technology and communication methods. 

For our \textbf{first example}, we take quantum phase estimation, a ubiquitous quantum procedure that is used in many other fundamental quantum algorithms. This presents a first opportunity to look at task modeling from a QC perspective. The phase estimation problem is formulated as follows: \textit{Given a unitary operator $U$ and eigenstate $\left|\Psi\right\rangle$ with an unknown eigenvalue $e^{i2\pi\varphi}$, estimate the value $\varphi$.}

Some of the assumptions on this problem come from the underlying physics and hardware, that $U$ (as well as any of its corresponding control operations) can be efficiently implemented by a quantum circuit, and that we have the ability to efficiently prepare the state $\left|\Psi\right\rangle$. Furthermore, we impose the mathematical assumption that (for simplicity) $\varphi$ can be written in an exact binary representation, 
\[
\varphi=0.\varphi_1\varphi_2\cdots\varphi_m.
\]
One way to construct $\varphi$ is by iteratively determining $\varphi_i$ from the least significant to the most significant bit. To illustrate a synchronous model, for each $\varphi_i$ we require two quantum bits for computation and a classical bit to store measured results. To determine the least significant bit $\varphi_m$, we (1) initialize the two qubits (one auxiliary, one in the eigenstate); (2) apply a certain operation (in this case, controlled $U^{2^{m-1}}$); (3) measure the auxiliary qubit in a certain basis (in this case, Pauli $X$); and (4) store the result in the classical register.

After this process is complete, the auxiliary qubit is reinitialized and then a \textit{phase correction} is performed in order to remove the contribution from $\varphi_m$ in order to identify the next significant bit, $\varphi_{m-1}$ in pretty much the same manner as described above (but now with the new phase), and so on and forth until $\varphi_1$.\ This is the notion of \textit{classical feedback}: the next stage of computation depends on the value of the classical register holding the previous result. This example relies on a synchronous execution paradigm, in which results from programs submitted to the QPU depend on successful execution of an ordered sequence of computations with the final result collected when available.

To contrast this, the $\varphi$ construction process can be executed in a \textit{parallel} fashion, where each precision bit is realized in its own sequence of initialization and control-$U$ operations with the results stored in its own classical register. This would require more qubits, larger circuit depth, and more classical registers to store the precision data. However, it is an illustration of asynchronous quantum tasks, introducing the potential to manage them at a higher level. The caveat here is that the synchronicity requirement is still imposed before the final measurement in the form of the inverse quantum Fourier transform that is applied to the results of the previous asynchronous steps. The inverse QFT is necessary for producing the correct transformation for estimating the phase. In this sense, the example would not be illustrative of a `fully asynchronous' process.

For our \textbf{second example}, we consider an algorithm that uses a quantum computer to create and measure the properties of a parameterized trial wave function, and a classical computer to optimize the wave-function parameters. In this case, error detection during the quantum computation could present a different kind of opportunity for synchronous task modeling within a hybrid quantum-classical computation~\cite{Urbanek2020}. (a) In the first stage of the quantum computation, the circuit would use an ancilla qubit, $a_1$, to detect an error in the preparation of the initial logical state. The detection of an error from the measurement of this ancilla qubit would reset the stage. (b) In the second stage of the quantum computation, a second ancilla qubit, $a_2$, is entangled with the logical qubit and is utilized to perform an arbitrary angle rotation; the result for its measurement indicates which angle $\theta$ was used in the computation. (c) From there, the expectation values (depending on $\theta$) are computed. The classical computation would then identify the angle that minimizes the algorithm's cost function. 

In principle, the actions on the two ancillas are independent of each other and can be performed asynchronously. However, the synchronous model applies when both the detection of an error in state preparation and the determination of the angle $\theta$ are combined into a single ancilla. In such a scenario, upon the right measurement outcome on $a_1$, this qubit is reset and the circuitry on $a_2$ follows on the same qubit.

Ultimately, noisy intermediate-scale quantum (NISQ) and fault-tolerant operation of QHPC systems will introduce separate requirements for asynchronous multi-tasking. In the NISQ context, we need to periodically characterize the quantum processors to estimate noise parameters and perform error mitigation to finalize QPU results. There is also the concern of making calls to multiple QPUs as they lack the reliability and reproducibility of traditional technology; this introduces additional programming concerns.

In the context of fault-tolerant quantum error correction (FTQEC), we need to process conditional statements that may affect the runtime of the QPU themselves. This will lead to uncertainty in device operation times and, hence, will be better treated as an asynchronous request. Moreover, the devices are inherently probabilistic and post-processing will require gathering statistical confidence that affects runtime. This leads to an intersection between requirements for QC and other probabilistic computing paradigms.

\section{Perspective on the Role of Quantum Technology}

Coupling HPC and QC systems will affect the way we interact with the latter. For instance, a QC system might only be accessible through a cloud environment, potentially leading to queue waiting and larger communication latency.  Conversely, a tighter coupling with the HPC system will result in more efficient execution due to the availability of specialized encoding and a tailored instruction set, but this will also require the runtime to be more responsive, thus increasing the computational load of the classical side.

A leading performance concern is the choice of a system-level/compiled programming model versus an interpreted (Python) programming model that is very common for QC today. We advocate for more integration at the system level, notably via libraries and their direct management of system resources. In addition, the examples described in Section~\ref{sec:examples} are relatively simple but illustrate that quantum data types and a standardizing format of data structures can lead to performance advantages, especially when leveraged to take advantage of classical resources. 

In task-based models for irregular, dynamic and heterogeneous applications, critical paths are probabilistic and we cannot estimate them a priori. Schedulers, resource managers, and runtimes will have to face a challenging transformation in order to minimize idle times and leverage asynchronous operations. Furthermore, the ability to manifest the probabilistic behavior expected from quantum computers is hindered by noise from the environment, and how this kind of additional complexity affects quantum-classical task management might reveal its importance in the future.

\section*{\footnotesize Acknowledgements}

\footnotesize{EW, DC, TSH, and ED acknowledge that this work was performed at Oak Ridge National Laboratory, operated by UT-Battelle, LLC under contract DE-AC05-00OR22725 for the US Department of Energy (DOE). EW, DC, and TSH acknowledge that support for the work came from the DOE Advanced Scientific Computing Research (ASCR) Accelerated Research in Quantum Computing (ARQC) Program under field work proposal ERKJ332. ED is supported by the DOE Office of Science Advanced Scientific Research Program Early Career Award under contract number 3ERKJ420. SLA acknowledges that this material is partially based upon work supported by the National Science Foundation under Award No. 2300476.}

%
%
\bibliographystyle{splncs04}
\bibliography{bibliography}

\begin{thebibliography}{10}
\providecommand{\url}[1]{\texttt{#1}}
\providecommand{\urlprefix}{URL }
\providecommand{\doi}[1]{https://doi.org/#1}

\bibitem{Arute2019}
Arute, F., {et al.}: Quantum supremacy using a programmable superconducting
  processor. Nature  \textbf{574},  505--510 (2019).
  \doi{10.1038/s41586-019-1666-5}

\bibitem{Britt2017}
Britt, K.A., Mohiyaddin, F.A., Humble, T.S.: Quantum accelerators for
  high-performance computing systems. In: 2017 IEEE International Conference on
  Rebooting Computing (ICRC). pp.~1--7 (2017). \doi{10.1109/ICRC.2017.8123664}

\bibitem{Chen2024}
Chen, C., Nguyen, D.T., Lee, S.J., Baker, N.A., Karakoti, A.S., Lauw, L., Owen,
  C., Mueller, K.T., Bilodeau, B.A., Murugesan, V., Troyer, M.: Accelerating
  computational materials discovery with artificial intelligence and cloud
  high-performance computing: from large-scale screening to experimental
  validation (2024). \doi{10.48550/arXiv.2401.04070}

\bibitem{LopezAlarcon2022}
{Lopez Alarcon}, S., Elster, A.C.: Quantum computing and high-performance
  computing: Compilation stack similarities. Computing in Science \&
  Engineering  \textbf{24}(06),  66--71 (Nov 2022).
  \doi{10.1109/MCSE.2023.3269645}

\bibitem{LopezAlarcon2023}
{Lopez Alarcon}, S., Wong, E., Humble, T., Dumitrescu, E.: Quantum programming
  paradigms and description languages. Computing in Science \& Engineering
  (2024), {To appear.}

\bibitem{Madsen2022}
Madsen, L.S., {et al.}: {Quantum Computational Advantage With a Programmable
  Photonic Processor}. Nature  \textbf{606},  75--81 (2022).
  \doi{10.1038/s41586-022-04725-x}

\bibitem{Mccaskey2020}
Mc{C}askey, A., Lyakh, D., Dumitrescu, E.F., Powers, S., Humble, T.: {XACC}: a
  system-level software infrastructure for heterogeneous quantum–classical
  computing. Quantum Science and Technology  \textbf{5}(2),  1--23 (2020).
  \doi{10.1088/2058-9565/ab6bf6}

\bibitem{Mintz2020}
Mintz, T.M., McCaskey, A.J., Dumitrescu, E.F., Moore, S.V., Powers, S.,
  Lougovski, P.: Qcor: A language extension specification for the heterogeneous
  quantum-classical model of computation. J. Emerg. Technol. Comput. Syst.
  \textbf{16}(2) (mar 2020). \doi{10.1145/3380964}

\bibitem{Schulz2022}
Schulz, M., Ruefenacht, M., Kranzlmuller, D., Schulz, L.: Accelerating hpc with
  quantum computing: It is a software challenge too. Computing in Science \&
  Engineering  \textbf{24}(04),  60--64 (jul 2022).
  \doi{10.1109/MCSE.2022.3221845}

\bibitem{Urbanek2020}
Urbanek, M., Nachman, B., de~Jong, W.A.: {Error Detection on Quantum Computers
  Improving the Accuracy of Chemical Calculations}. Phys. Rev. A  \textbf{102},
   022427 (Aug 2020). \doi{10.1103/PhysRevA.102.022427}

\end{thebibliography}

\end{document}